\documentclass[twocolumn,superscriptaddress,amsmath,amssymb]{revtex4-1}
\usepackage{graphicx,bm,dsfont}
\usepackage[colorlinks=true,citecolor=blue,urlcolor=blue]{hyperref}

\begin{document}
\title{Collectively induced transparency and absorption in waveguide QED with Bragg atom arrays}
\author{Haolei Cheng}
\affiliation{Center for Joint Quantum Studies and Department of Physics, School of Science, Tianjin University, Tianjin 300350, China}
\author{Wei Nie}\email{weinie@tju.edu.cn}
\affiliation{Center for Joint Quantum Studies and Department of Physics, School of Science, Tianjin University, Tianjin 300350, China}
\affiliation{Tianjin Key Laboratory of Low Dimensional Materials Physics and Preparing Technology, Tianjin University, Tianjin 300350, China}

\begin{abstract}
Collective quantum states, such as subradiant and superradiant states, are useful for controlling optical responses in many-body quantum systems. In this work, we study novel collective quantum phenomena in waveguide-coupled Bragg atom arrays with inhomogeneous frequencies. For atoms without free-space dissipation, collectively induced transparency is produced by destructive quantum interference between subradiant and superradiant states. In a large Bragg atom array, multi-frequency photon transparency can be obtained by considering atoms with different frequencies. Interestingly, we find collectively induced absorption (CIA) by studying the influence of free-space dissipation on photon transport. Tunable atomic frequencies nontrivially modify decay rates of subradiant states. When the decay rate of a subradiant state equals to the free-space dissipation, photon absorption can reach a limit at a certain frequency. In other words, photon absorption is enhanced with low free-space dissipation, distinct from previous photon detection schemes. We also show multi-frequency CIA by properly adjusting atomic frequencies. Our work presents a way to manipulate collective quantum states and exotic optical properties in waveguide QED systems.
\end{abstract}

\maketitle

\section{Introduction}
One-dimensional (1D) waveguides provide an important interface for light-matter interaction~\cite{lodahl2017chiral,RevModPhys.90.031002,RevModPhys.95.015002,gonzalez2024light}. Waveguide quantum electrodynamics (QED) has broad applications in quantum devices, including optical switches~\cite{PhysRevLett.95.213001,shen2005coherent,chang2007single,PhysRevLett.101.100501,PhysRevA.80.062109,PhysRevA.81.042304,PhysRevA.83.023811,PhysRevA.84.045801,PhysRevA.89.053813,PhysRevA.96.053832,Jia2022,
zheng2023few}, photon detectors~\cite{PhysRevLett.102.173602,inomata2016single,PhysRevX.10.041054,PhysRevApplied.15.044041}, quantum storage~\cite{PhysRevX.7.031024,PhysRevLett.127.250402,zanner2022coherent} and quantum networks~\cite{PhysRevX.7.011035,PhysRevLett.118.133601,PhysRevApplied.15.054043}. Recently, multi-atom waveguide QED systems receive growing attention in theory~\cite{PhysRevA.88.043806,PhysRevA.92.053834,PhysRevA.94.043844,PhysRevA.94.053842,PhysRevA.95.033818,PhysRevA.96.013842,PhysRevA.96.033857,PhysRevA.96.043872,PhysRevA.98.023814,PhysRevLett.124.093604,PhysRevA.102.033728,
PhysRevLett.129.253601,qiu2023non} and experiment~\cite{VanLoo2013,PhysRevLett.115.063601,solano2017super,mirhosseini2019cavity,PhysRevLett.123.233602,Kannan2020,tiranov2023collective}. For Bragg atom arrays, i.e., the distance between nearest-neighboring atoms is $n\lambda_0/2$ ($\lambda_0$ is the single-photon wavelength corresponding to atomic frequency, $n$ is an integer), photon-mediated coherent couplings are vanishing. This gives rise to a superradiant state with decay rate $N\Gamma$, where $N$ is the atom number in the array and $\Gamma$ is the single-atom decay rate. As a consequence, light reflection is enhanced, making Bragg atom arrays useful for quantum mirrors~\cite{chang2012cavity,PhysRevLett.117.133603,PhysRevLett.117.133604,song2021optical}. However, many quantum optical effects in waveguide QED systems arise from quantum interference between superradiant and subradiant states~\cite{RevModPhys.95.015002,PhysRevLett.131.103602}. Therefore, it is critical to manipulate collective quantum states for controlling photon transport in waveguide QED with Bragg atom arrays.

Cavity QED with many atoms provides an alternative way to study collective quantum phenomena~\cite{PhysRevLett.119.093601,PhysRevLett.124.023603,PhysRevLett.126.123602,PhysRevLett.127.073603,PhysRevX.12.011054}. In the limit of bad cavity, cavity-coupled atoms show collective behaviors~\cite{mlynek2014}. Recently, a novel cooperative quantum phenomenon, i.e., collectively induced transparency (CIT), is studied in a cavity with inhomogeneous atoms~\cite{Lei2023}. In the bad-cavity regime, atoms with different frequencies produce subradiant states. Strong driving of the system can efficiently excite subradiant atomic states, and gives rise to narrow transparency windows in the cavity transmission spectrum. Simply speaking, CIT in cavity QED is caused by quantum interference between optical responses of collective quantum states. In cavity QED, atomic dissipation is induced by lossy cavity. The role played by other dissipation channels in CIT has not been studied.

\begin{figure}[b]
\includegraphics[width=8cm]{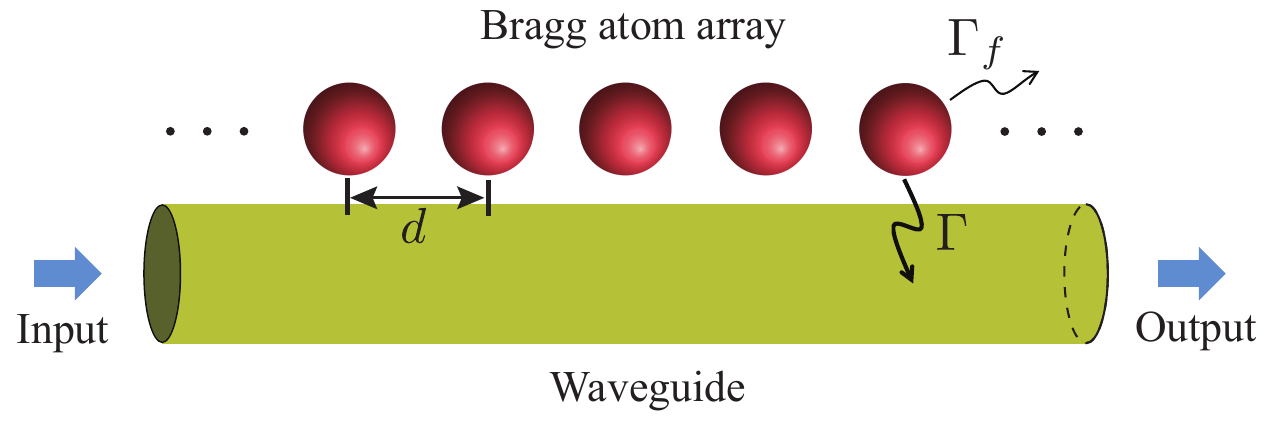}
\caption{Schematic of waveguide QED with a Bragg atom array. Atoms in the array are two-level systems. Nearest-neighboring atoms have homogeneous spacing $d=n\lambda_0/2$ with an integer $n$. We assume that all the atoms have decay rates $\Gamma$ and $\Gamma_f$, due to couplings to the waveguide and free space, respectively.}\label{fig1}
\end{figure}

In this work, we study photon transport in a waveguide coupled to a Bragg atom array. We show that atoms with different frequencies give rise to CIT in the waveguide, similar to its counterpart in cavity QED~\cite{Lei2023}. To understand this effect, we focus on photon transparency in a two-atom system. Two atoms with the same frequency produce a superradiant state and a dark state. The two-atom system is protected by anti-$\mathcal{PT}$ symmetry. Weak frequency shift of a single atom changes the dark state to a subradiant state with weak dissipation. In the transmission process, destructive quantum interference takes place between photons scattered by subradiant and superradiant states, yielding CIT. We generalize our discussion to Bragg array with $N$ atoms. When multiple atoms have different frequencies, there is multi-frequency photon transparency. More importantly, we find that CIT changes to collectively induced absorption (CIA) when free-space dissipation of atoms is considered. The CIA effect is realized when the free-space dissipation equals to decay rate of a subradiant state. For Bragg atom arrays with many detuned atoms, it is possible to realize multi-frequency CIA. Our work presents novel features of collective quantum phenomena in waveguide QED.

\section{Collectively induced transparency in waveguide QED}
Photon transport in a waveguide is modified by the spatial structure of atom array. Under Bragg scattering condition, the atom array behaves as a superatom with a single superradiant state. Incident photons are completely reflected by the superradiant state. Here, we study how inhomogeneous atomic frequencies alter photon transport in the waveguide.

\subsection{Theory of single-photon transport in a waveguide}
As shown in Fig.~\ref{fig1}, we study an array with $N$ atoms coupled to a 1D waveguide. In the Born-Markov approximation, dynamics of atoms in the waveguide can be described by a master equation ($\hbar=1$)
\begin{equation}
\dot{\rho}(t)=-i[H_{0}+H_{\mathrm{W}},\rho(t) ]+\mathcal{D}[\rho],
\end{equation}
with $H_{0}=\omega_{0}\sum_{i}\sigma_{i}^{+}\sigma _{i}^{-}$. Operators of atoms are $\sigma_{i}^{+}= \left | e_{i} \right\rangle  \left\langle g_{i} \right|$ and $\sigma_{i}^{-}= \left | g_{i} \right\rangle  \left\langle e_{i} \right|$, where $\left | g_{i} \right\rangle$ and $\left | e_{i} \right\rangle$ are the ground state and excited state of the $i$th atom, respectively. Photon-mediated coherent couplings are described by the Hamiltonian $H_{\mathrm{W}}=\sum_{ij}g_{ij}\sigma_{i}^{+}\sigma_{j}^{-}$, with $g_{ij}=\sqrt{\Gamma _{i}\Gamma_{j}}\sin (2\pi d_{ij}/\lambda _{0})$. Here, $d_{ij}$ denotes the distance between $i$th and $j$th atoms, and $\lambda_0$ is the single-photon wavelength corresponding to the frequency $\omega_0$. For the Bragg scattering condition, i.e., $d_{ii+1}=n\lambda_0/2$, coherent couplings are vanishing. In addition to coherent atom-atom interactions, continuous photonic modes in the waveguide give rise to correlated dissipations, or dissipative couplings, between atoms. This is described by the Lindblad operator

\begin{equation}
D[\rho]=\sum_{i,j}\gamma _{ij}(2\sigma_{i}^{-}\rho\sigma_{j}^{+}-\sigma_{i}^{+}\sigma_{j}^{-}\rho-\rho\sigma_{i}^{+}\sigma_{j}^{-}),
\end{equation}
with $\gamma_{ij}=\sqrt{\Gamma _{i}\Gamma_{j}}\cos (2\pi d_{ij}/\lambda _{0})$. Without loss of generality, we consider homogeneous atomic dissipation $\Gamma$ induced by the waveguide. Therefore, the effective Hamiltonian of the Bragg atom array system with $d=\lambda_{0}/2$ between nearest-neighboring atoms is
\begin{equation}
H_{\mathrm{eff}}  =  { \sum_{i}} (\omega_{0}-i\Gamma)\sigma _{i}^{+}\sigma _{i}^{-}-i\Gamma{ \sum_{i\neq j}} (-1)^{|i-j|}\sigma _{i}^{+}\sigma _{j}^{-}.
\end{equation}
In the single-excitation subspace, we diagonalize the non-Hermitian Hamiltonian $H_{\mathrm{eff}}=\sum_j E_j |\psi_j^R\rangle \langle \psi_j^L|$, with the biorthogonal basis $\langle \psi_j^L|\psi_{j'}^R \rangle=\delta_{jj'}$. We can find that $\mathrm{Re}(E_j)=\omega_0$. There is a superradiant state with decay rate $N\Gamma$. The remaining states have vanishing decay rate, i.e., dark states. In terms of these eigenmodes, the amplitudes of single-photon transmission and reflection are~\cite{PhysRevA.92.053834,PhysRevApplied.15.044041}
\begin{eqnarray}\label{Eqt}
t(\omega) &=&  1-i\Gamma \sum_{j}\frac{\Xi_j}{\omega-E_{j}},\label{Eqt} \\
r(\omega) & = & -i\Gamma \sum_{j} \frac{\tilde{\Xi}_j}{\omega-E_{j}},\label{Eqr}
\end{eqnarray}
with the frequency $\omega$ of incident photon, and interaction spectra $\Xi_j = V^{\dag}|\psi_{j}^R\rangle \langle\psi_{j}^L| V $ and $\tilde{\Xi}_j = V^{\top}|\psi_{j}^R\rangle \langle\psi_{j}^L| V $ for transmission and reflection processes, respectively. Here, $V = (e^{ik_{0}x_{1}},e^{ik_{0}x_{2}},\cdots)$, where $k_0=\omega_0/c$, and $x_i$ denotes the position of the $i$th atom in the waveguide. The interaction spectra $\Xi_j$ and $\tilde{\Xi}_j$ characterize overlaps of the propagating photonic mode and eigenmodes of the effective Hamiltonian in scattering processes. For the Bragg atom array with homogeneous frequency, the interaction spectrum of the superradiant state is $\Xi = N$, and the interaction spectra of dark states are zero. According to Eq.~(\ref{Eqt}), incident photon cannot transmit through the atom array at resonance. However, if atoms have different frequencies, subradiant states can be produced. Therefore, photon transport in the waveguide is determined by optical responses of subradiant and superradiant states.

\subsection{A minimal model: two-atom system}
\begin{figure}[t]
\includegraphics[width=8.5cm]{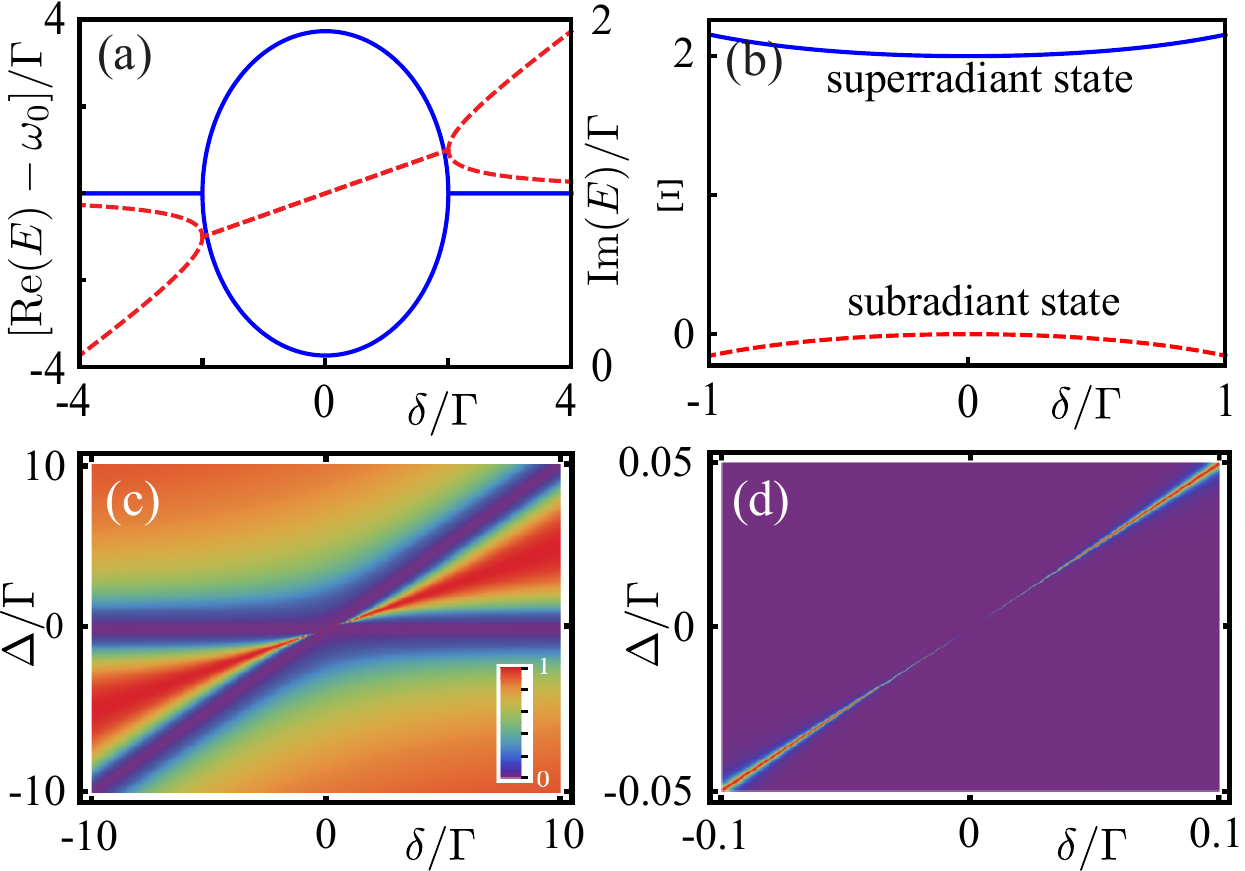}
\caption{(a) Real (blue-solid) and imaginary (red-dashed) parts of eigenspectrum in the two-atom system. Two degenerate subradiant and superradiant states are formed in the parameter regime $|\delta|<2\Gamma$. (b) Interaction spectrum $\Xi$ for subradiant and superradiant states. (c) Photon transmission induced by frequency detuning between two atoms. (d) CIT for weak frequency detuning.}\label{fig2}
\end{figure}

We consider two atoms coupled to a waveguide in the Bragg scattering condition. The frequency of an atom is shifted by $\delta$. The frequency detuning $\delta$ is so small comparing to atomic frequency $\omega_0$ that photon-mediated dissipative coupling between two atoms is not changed. The effective non-Hermitian Hamiltonian of the two-atom system is
\begin{eqnarray}\label{EqH2}
H_{\mathrm{eff},2}& =&  (\omega_{0}+\delta-i\Gamma)\sigma _{1}^{+}\sigma _{1}^{-}+ (\omega_{0}-i\Gamma) \sigma _{2}^{+}\sigma _{2}^{-} \nonumber \\
&& + i\Gamma(\sigma _{1}^{+}\sigma _{2}^{-}+\sigma _{2}^{+}\sigma _{1}^{-} ).
\end{eqnarray}
In the single-excitation subspace, the Hamiltonian can be written as
\begin{eqnarray}
H_{\mathrm{eff},2} = \begin{pmatrix}  \omega_{0}+\delta-i\Gamma&i\Gamma
 \\  i\Gamma&\omega_{0}-i\Gamma  \end{pmatrix}.
\end{eqnarray}
We define $\tilde{H}_{\mathrm{eff},2}=H_{\mathrm{eff},2}-(\omega_0+\delta/2) \bm{\mathds{1}}$. Due to waveguide-mediated dissipative coupling, the Hamiltonian $\tilde{H}_{\mathrm{eff},2}$ is protected by anti-parity-time (anti-$\mathcal{PT}$) symmetry~\cite{peng2016anti,PhysRevLett.126.180401},
\begin{equation}
\mathcal{P T}\tilde{H}_{\mathrm{eff},2}(\mathcal{P T})^{-1}=-\tilde{H}_{\mathrm{eff},2}.
\end{equation}
Eigenvalues of the non-Hermitian Hamiltonian $H_{\mathrm{eff},2}$ are
\begin{eqnarray}
E_{1}&=&\omega_{0}-i\Gamma+\frac{\delta +\mu}{2}, \\
E_{2}&=&\omega_{0}-i\Gamma+\frac{\delta -\mu}{2},
\end{eqnarray}
with $\mu=\sqrt{\delta^{2}-4\Gamma^{2}}$. For $-2\Gamma<\delta<2\Gamma$, real parts of $E_1$ and $E_2$ are the same. However, their imaginary parts are different, giving rise to subradiant and superradiant states. Hence, these two collective states have degenerate energy levels. For $\delta<-2\Gamma$ or $\delta>2\Gamma$, real parts of $E_1$ and $E_2$ become different, but their imaginary parts are the same. The points at $\delta=\pm 2\Gamma$ correspond to exceptional points. In Fig.~\ref{fig2}(a), we show the real and imaginary parts of these two states. At $\delta=0$, the eigenvalues of the Hamiltonian are $E_{1}=\omega_{0}$ and $E_{2}=\omega_{0}-2i\Gamma$, corresponding to a dark state and a superradiant state, respectively. Weak frequency shift produces a subradiant state. The right vectors of these two states are
\begin{eqnarray}
\left | \psi_{1}^R  \right \rangle &=& \frac{-i}{\sqrt{\mathcal{N}_1}} \Big(\frac{\delta+\mu}{2\Gamma},i\Big)^{\top},
\\\left | \psi_{2}^R  \right \rangle &=&\frac{-i}{\sqrt{\mathcal{N}_2}} \Big (\frac{\delta-\mu}{2\Gamma},i\Big)^{\top},
\end{eqnarray}
with normalization factors $\mathcal{N}_{1} = 2\mu/(\mu-\delta)$ and $\mathcal{N}_{2} = 2\mu/(\mu+\delta)$. Transposes of these right vectors become left vectors of corresponding collective states. At $\delta=0$, these two eigenstates are Bell states: $|\psi_{1} \rangle =\frac{1 }{\sqrt{2}} (1,-1)^{\top}$ and $| \psi_{2}\rangle =\frac{1 }{\sqrt{2}} (1,1)^{\top}$.

Interaction spectra for the subrradiant and superradiant states are
\begin{eqnarray}
\Xi_1 &=& \frac{\mu - i2\Gamma}{\mu}, \\
\Xi_2 &=& \frac{\mu + i2\Gamma}{\mu},
\end{eqnarray}
respectively. The photon transmission amplitude becomes
\begin{equation}\label{Eqt2atom}
t = 1- i\Gamma \left(\frac{\Xi_1}{\Delta-\frac{\delta+\mu}{2} + i\Gamma} + \frac{\Xi_2}{\Delta-\frac{\delta-\mu}{2}+i\Gamma} \right),
\end{equation}
with the detuning $\Delta = \omega - \omega_0$ between the frequency of incident photon and atomic frequency. In Fig.~\ref{fig2}(b), we show the interaction spectra $\Xi_1$ and $\Xi_2$ for subradiant and superradiant states at weak atomic frequency shift $\delta$. At $\delta=0$, $\Xi_1=0$ and $\Xi_2=2$. Atomic frequency shift nontrivially modifies interaction spectra, as well as photon transport in the waveguide. The transmission amplitude in Eq.~(\ref{Eqt2atom}) can be simplified as
\begin{equation}
t= \frac{\Delta (\Delta-\delta)}{\Delta(\Delta-\delta)+ i \Gamma(2\Delta-\delta)}.
\end{equation}
The photon transmission defined as $T=|t|^2$ vanishes at $\Delta=0$ and $\Delta=\delta$, as shown in Fig.~\ref{fig2}(c). In fact, the vanishing transmission at $\delta=0$ is produced by the superradiant state. If a single-atom frequency shift $\delta$ is weak, there is a narrow transmission peak. For clarity, we show transmission spectrum of the system for a small value of $\delta$ in Fig.~\ref{fig2}(d). From above equation, we can obtain the condition for photon transparency $\Delta=\delta/2$, i.e., incident photon should be resonant with the subradiant state.  Because of opposite phases for $\Xi_1$ and $\Xi_2$, destructive quantum interference takes place in photon transmission process. At $\Delta=\delta/2$, the subradiant and superradiant states completely cancel each other in the optical response, leading to photon transparency, i.e., CIT~\cite{Lei2023}.

\subsection{Waveguide QED with Bragg atom arrays}
\begin{figure}[t]
\includegraphics[width=8.5cm]{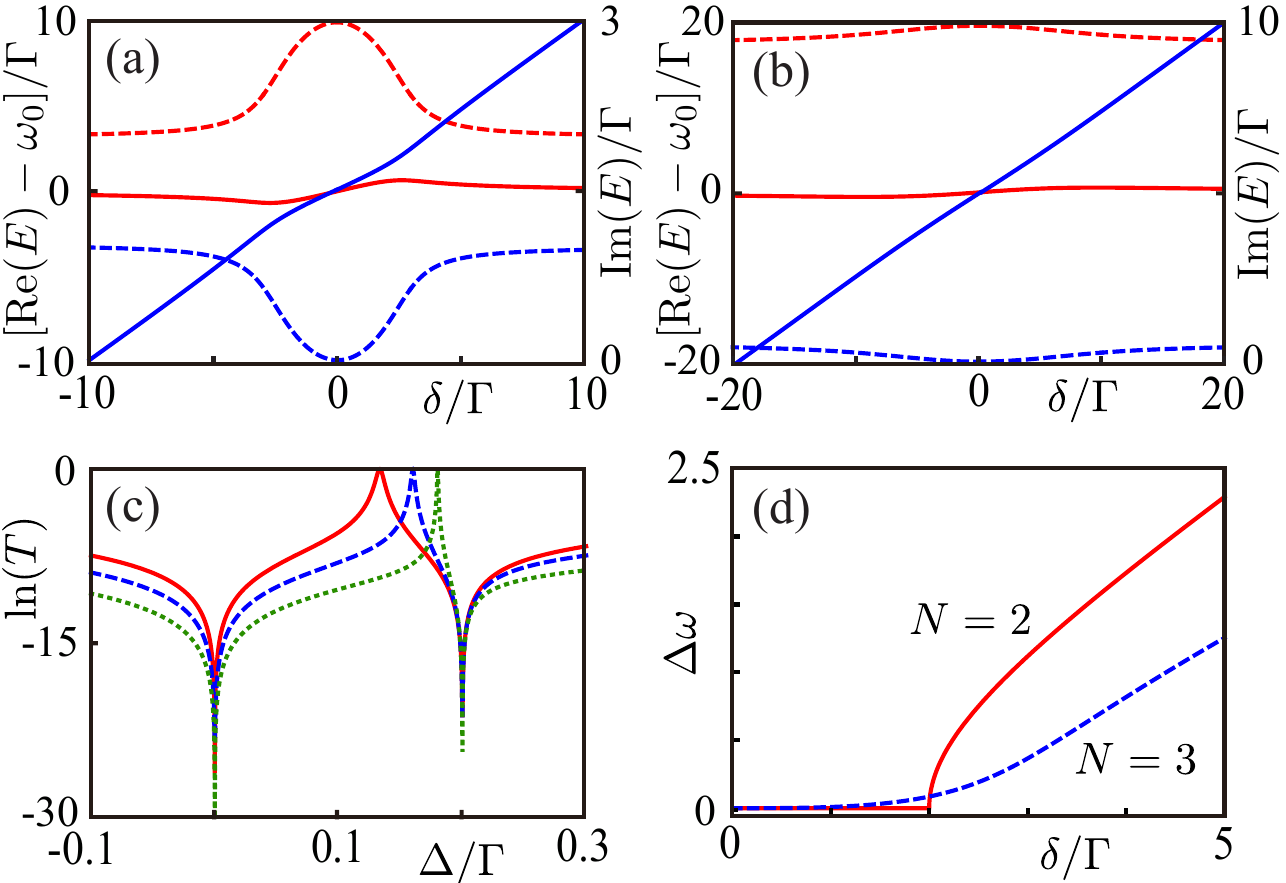}
\caption{(a) Energy levels (solid) and decay rates (dashed) of superradiant (red) and subradiant (blue) states in a three-atom system. (b) Energy levels and decay rates of superradiant and subradiant states in the Bragg array with $10$ atoms. (c) Logarithm of photon transmission versus $\Delta$ in systems with $N=3$ (red-solid), $N=5$ (blue-dashed) and $N=10$ (green-dotted). (d) Frequency difference $\Delta \omega$ between transparency window and subradiant state.}\label{fig3}
\end{figure}

We can generalize the system to a Bragg array with $N$ atoms. The effective Hamiltonian is
\begin{equation}
H_{\mathrm{eff},N}  =  { \sum_{i}} (\omega_{0}+\delta_i-i\Gamma)\sigma _{i}^{+}\sigma _{i}^{-} -i\Gamma{ \sum_{i\neq j}} (-1)^{|i-j|}\sigma _{i}^{+}\sigma _{j}^{-}.
\end{equation}
Without loss of generality, we assume that the frequency of the first atom is shifted, i.e., $\delta_1=\delta, \delta_{i\neq1}=0$. There are two collective states which depend on the single-atom frequency shift $\delta$. Their eigenvalues are
\begin{eqnarray}
E_{1} &=&\omega_{0}+\frac{\delta }{2}-\frac{iN\Gamma-\mu}{2}, \\
E_{2}&=&\omega_{0}+\frac{\delta }{2}-\frac{iN\Gamma+\mu}{2},
\end{eqnarray}
with
\begin{eqnarray}
\mu =\sqrt{-N^{2}\Gamma^{2}+2(N-2)i\Gamma\delta+\delta^{2}}.
\end{eqnarray}
These two states are subradiant and superradiant states in the Bragg atom array. Other eigenstates are dark states which do not contribute to the photon transmission. In Figs.~\ref{fig3}(a) and \ref{fig3}(b), we show the energy levels and decay rates of two collective modes in systems with $N=3$ and $N=10$, respectively. There are no exceptional points in spectra of Bragg atom arrays for $N\geq 3$. A subradiant state is produced by the single-atom frequency shift. For the system with a large number of atoms, the decay rate and energy level of the superradiant state are slightly changed by $\delta$. However, the energy level of the subradiant state linearly depends on the single-atom frequency shift.

Right and left vectors of these two collective states are
\begin{eqnarray}
|\psi_{k}^R\rangle &=&\frac{(-1)^N}{\sqrt{\mathcal{N}_k}} \Big(\alpha_k, 1,-1,\cdots, (-1)^{N-2}\Big)^{\top}, \\
\langle\psi_{k}^L| &=& \frac{(-1)^N}{\sqrt{\mathcal{N}_k}} \Big(i\frac{\Gamma \mathcal{N}_k}{\mu}, 1,-1,\cdots, (-1)^{N-2}\Big),
\end{eqnarray}
with $\alpha_k=[(N-2)\Gamma-i\delta+i(-1)^k\mu]/2\Gamma$ and $\mathcal{N}_k = 2(N-1)\mu/[\mu+(-1)^k\delta+i(-1)^k (N-2)\Gamma]$ for $k=1,2$. Interaction spectra for the subradiant and superradiant states are
\begin{eqnarray}
\Xi_1 &=& \frac{\mu N-(N-2)\delta -iN^2\Gamma}{2\mu}, \\
\Xi_2 &=& \frac{\mu N+(N-2)\delta +iN^2\Gamma}{2\mu},
\end{eqnarray}
respectively. With these two interaction spectra, we can analytically derive the photon transmission amplitude
\begin{equation}
t = \frac{\Delta (\Delta - \delta)}{\Delta (\Delta - \delta) + i \Gamma(N \Delta - (N-1)\delta)}.
\end{equation}
In Fig.~\ref{fig3}(c), we show logarithm of photon transmission versus the detuning $\Delta$. As same as the two-atom system, the transmission vanishes at $\Delta=0$ and $\Delta=\delta$. The photon transparency appears at
\begin{equation}\label{EqTran}
\Delta = \Big(1-\frac{1}{N}\Big)\delta,
\end{equation}
which only depends on the single-atom frequency shift and the size of the system. The dependence of transparency window on $N$ explicitly shows the cooperative feature of CIT. In Fig.~\ref{fig3}(d), we show the difference $\Delta\omega$ between the frequency of transparency window and the frequency of subradiant state. For $N=2$, the transparency window coincides with the frequency of subradiant state in the regime with anti-$\mathcal{PT}$ symmetry protection, i.e., $|\delta|<2 \Gamma$. However, for systems $N\geq 3$, the absence of anti-$\mathcal{PT}$ symmetry breaks the frequency degeneracy in subradiant and superradiant states. Therefore, the transparency window is different from the frequency of subradiant state for $\delta\neq 0$.

\begin{figure}[b]
\includegraphics[width=8.5cm]{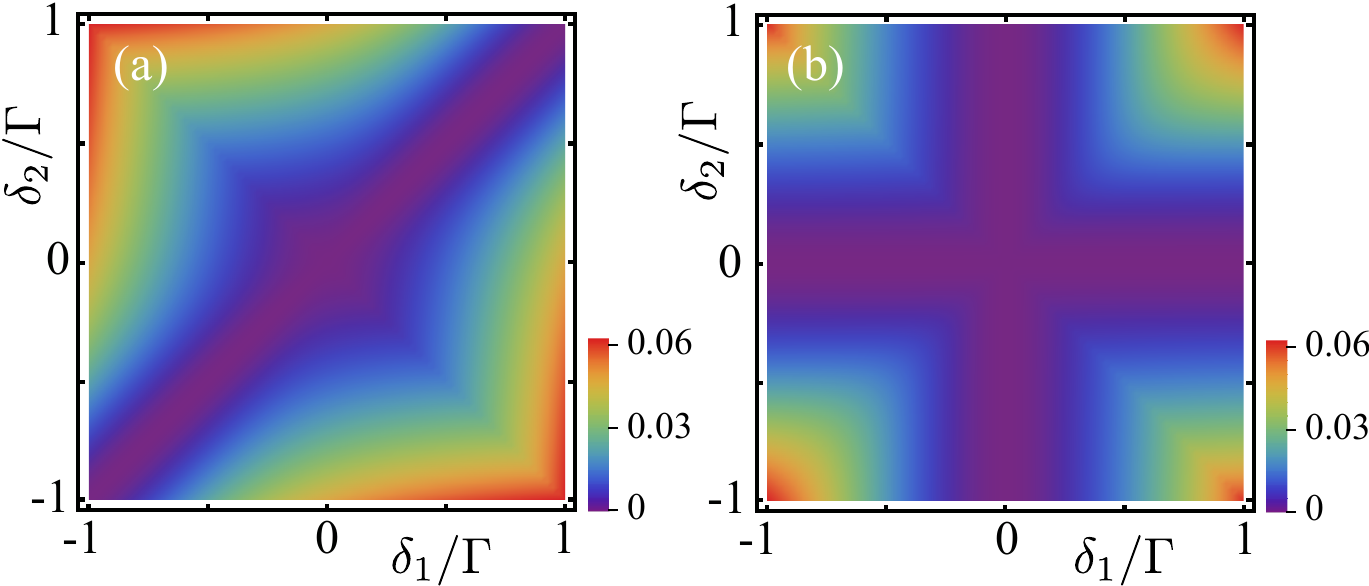}
\caption{(a),(b) Decay rates of two subradiant states. We consider a Bragg array with four atoms, where two atoms have frequency shifts $\delta_1$ and $\delta_2$, respectively.}\label{fig4}
\end{figure}

\subsection{Multiple inhomogeneous atoms}
From above discussions, we know that frequency shift of a single atom produces a subradiant state, which is responsible for a transparency window. Multi-window transmission spectrum can be obtained by shifting more atomic frequencies. As shown in Figs.~\ref{fig4}(a) and \ref{fig4}(b), decay rates of two subradiant states are changed by two frequency shifts of two atoms in the system with $N=4$. These two subradiant states are changed by frequency shifts in different ways. The subradiant state in Fig.~\ref{fig4}(a) has vanishing decay rate when two atoms have the same frequency shift. The subradiant state in Fig.~\ref{fig4}(b) has vanishing decay rate when only one atom is shifted. Therefore, by controlling frequencies of two atoms, we can modify decay rates of two subradiant states. In order to produce two subradiant states with nonnegligible decay rates, we should choose $\delta_1=-\delta_2$.

\begin{figure}[t]
\includegraphics[width=8.5cm]{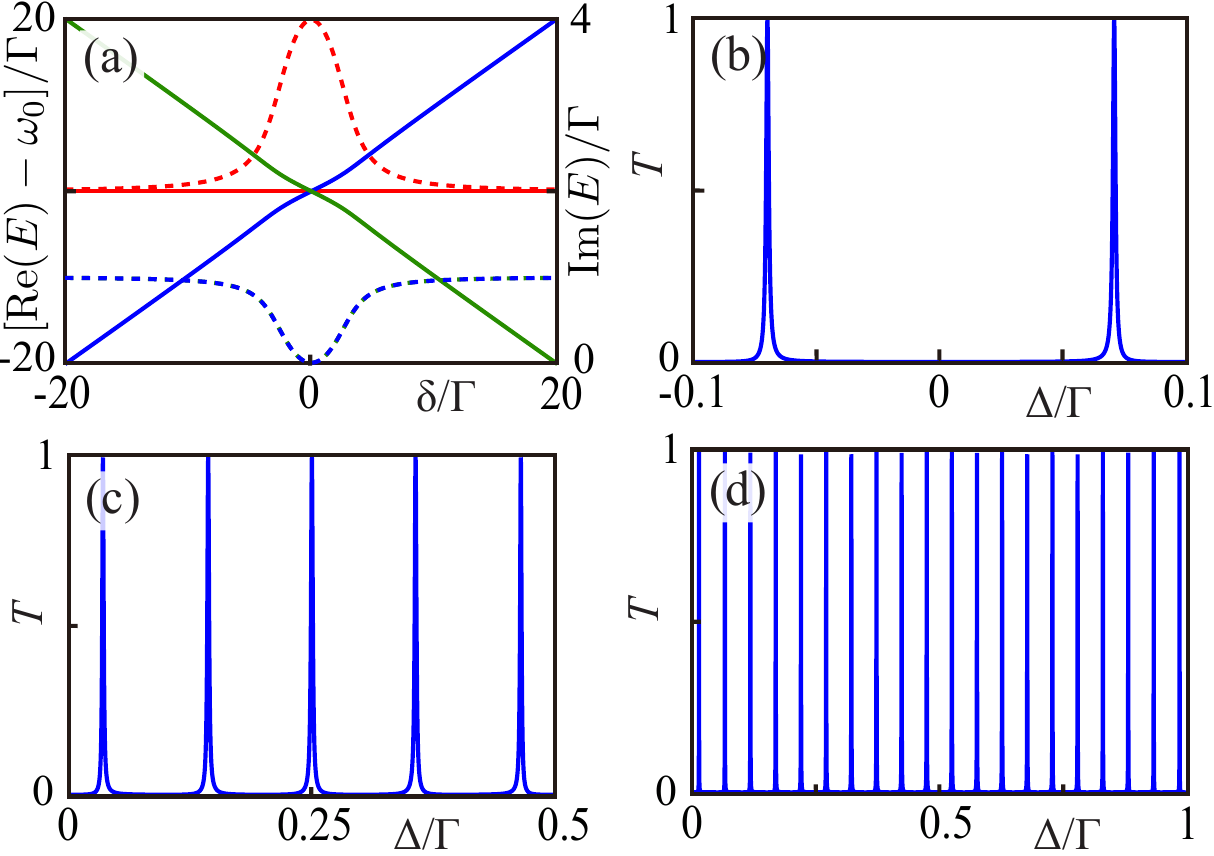}
\caption{(a) Energy levels and decay rates of collective states for Bragg atom array with $N=4$. Red-solid (-dashed), blue-solid (-dashed) and green-solid (-dashed) curves correspond to frequencies (decay rates) of the superradiant state and two subradiant states, respectively. Here, we consider that two atoms have frequency shifts $\delta$ and $-\delta$, respectively. (b) Two-window CIT at $\delta=0.1\Gamma$. (c) Multi-window CIT for the Bragg atom array with $N=6$, which has frequency shifts with equal difference $\delta=0.1\Gamma$. (d) Transmission spectrum of the Bragg atom array with $N=21$, which has frequency shifts with equal difference $\delta=0.05\Gamma$.}\label{fig5}
\end{figure}

In Fig.~\ref{fig5}(a), energy levels and decay rates of subradiant and superradiant states vary with $\delta$ in the system with four atoms. We consider that two atoms have frequency shifts $\delta_1=-\delta_2=\delta$. There are two subradiant states with the same decay rates, but different energy levels. Energy level of the superradiant state is independent of frequency shifts, but its decay rate is modified. The transmission spectrum for a weak frequency shift is shown in Fig.~\ref{fig5}(b). There are two transparency windows corresponding to two subradiant states. Therefore, to achieve photon transmission with different frequencies, we should generate multiple subradiant states by tuning atomic frequencies. Here, we consider frequency shifts $(0,\delta,2\delta,\ldots,(N-1)\delta)$ with equal difference $\delta$. Figure~\ref{fig5}(c) shows multi-frequency photon transparency for a Bragg atom array which has frequency shifts with equal difference. When the size of the system is large, we find a transmission spectrum analogue to frequency comb, as shown in Fig.~\ref{fig5}(d).

\section{Collectively induced absorption}
\begin{figure}[t]
\includegraphics[width=8.5cm]{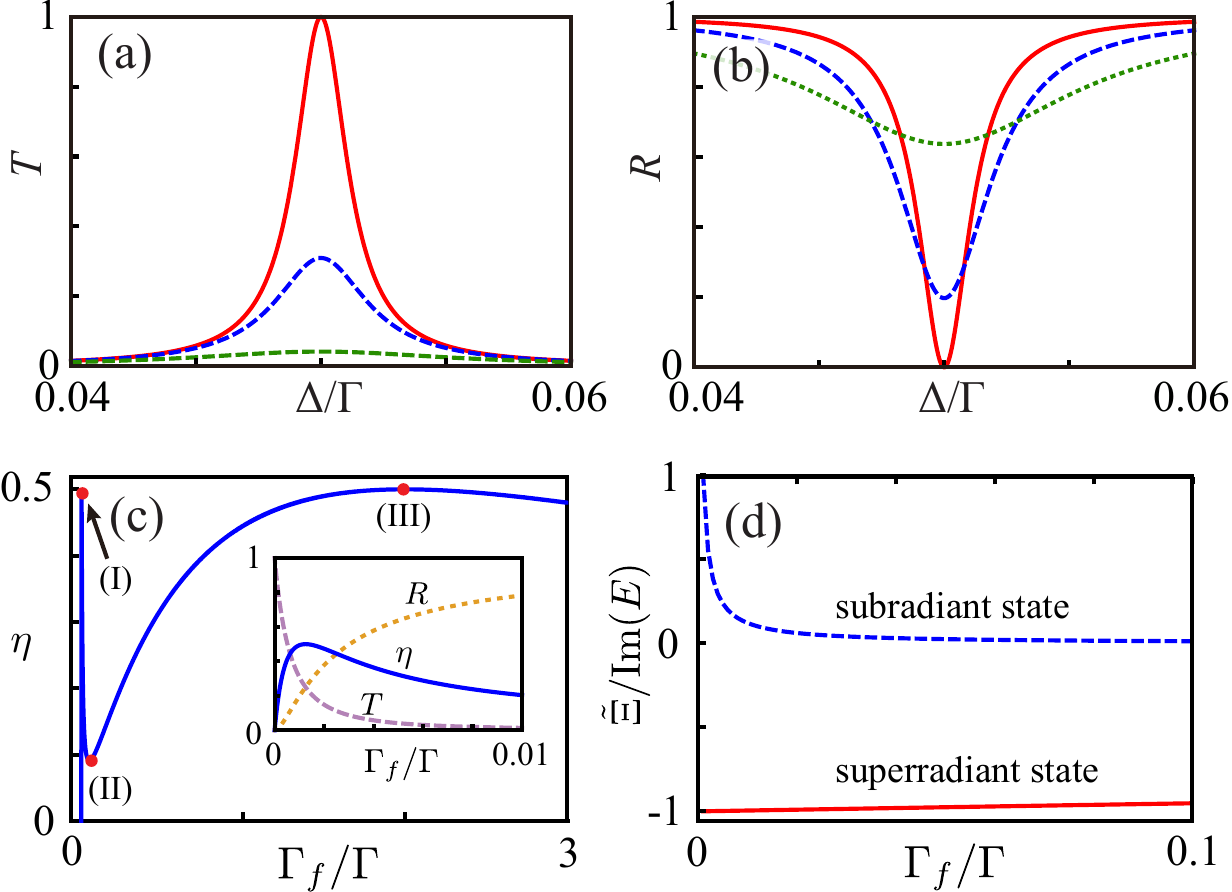}
\caption{(a) Transmission and (b) reflection for different values of free space dissipation. Red-solid, blue-dashed and green-dotted curves correspond to $\Gamma_f=0$, $0.001\Gamma$ and $0.005\Gamma$, respectively. (c) Photon absorption $\eta$ versus free-space dissipation. There are three extremal points. The inset shows transmission, reflection and absorption for weak free-space dissipation $\Gamma_f$. (d) Interaction spectra of collective states balanced by their decay rates in the reflection process. Here, we consider $N=2$ and $\delta=0.1\Gamma$.}\label{fig6}
\end{figure}

Up to now, we assume that the dissipation to free space, i.e., quantum channels except the waveguide, is negligible. In waveguide QED with superconducting artificial atoms, couplings between qubits and photonic modes in the waveguide are strong. Waveguide-induced dissipation plays a dominative role in controlling dynamics of qubits. Therefore, free-space dissipation is not taken into account in many studies. Here, we find that free-space dissipation is important for CIT. In Fig.~\ref{fig6}(a), we show the influence of weak free-space dissipation on photon transmission for the two-atom system. Photon transmission is reduced rapidly as the free-space dissipation $\Gamma_f$ increases. We also show photon reflection in Fig.~\ref{fig6}(b). Different from transmission process, photon reflection at resonance increases with $\Gamma_f$. So, free-space dissipation is important for photon transport controlled by subradiant states in waveguide QED. To evaluate photon transport in the waveguide, we define photon absorption $\eta = 1- |t|^2 -|r|^2$.

From Eqs.~(\ref{Eqt}) and (\ref{Eqr}), we can derive photon absorption at resonance as
\begin{eqnarray}\label{Eqeta}
\eta =  \frac{16\Gamma\Gamma_{f}(4\Gamma_{f}^{2}+\delta^{2})}{ (4\Gamma_{f}^{2}+8\Gamma \Gamma_{f}+\delta^{2})^{2} },
\end{eqnarray}
for the two-atom system. In Fig.~\ref{fig6}(c), we show the photon absorption is nontrivially changed by the free-space dissipation. Interestingly, photon absorption can reach the maximum for $\Gamma_0\ll \Gamma$, i.e., the point (I). The minimal photon absorption is obtained at point (II). Moreover, the maximal photon absorption can be also found with large free-space dissipation at point (III). To calculate values of free-space dissipation for these extremal photon absorption phenomena, we solve the equation $d\eta/d\Gamma_f=0$, and find three solutions:
\begin{eqnarray}
\Gamma_f &=& \Gamma-\frac{\sqrt{4\Gamma^{2}-\delta^{2}}}{2}, \\
\Gamma_f &=&\frac{\delta}{2}, \\
\Gamma_f &=& \Gamma+\frac{\sqrt{4\Gamma^{2}-\delta^{2}}}{2},
\end{eqnarray}
corresponding to $\mathrm{(I)}$, $\mathrm{(II)}$ and $\mathrm{(III)}$, respectively. These three cases represent interesting photon transport phenomena tuned by free-space dissipation. From our discussions about CIT, we know that subradiant and superradiant states play important roles in manipulating photon transmission. Therefore, it is critical to pinpoint the relation between collective atomic states in photon absorption. We discuss separately for these three cases.

In case (I), photon absorption is enhanced when the free-space dissipation equals to the decay rate of the subradiant state. The limit for photon absorption is $0.5$, as same as a single-atom photon detector~\cite{PhysRevLett.102.173602}. A single atom can reach the photon absorption limit when the free-space dissipation equals to atomic decay rate to waveguide. This means the atom should have large free-space dissipation, which constrains quantum coherence of the atom. Here, the condition for enhanced photon absorption is that the free-space dissipation equals to decay rate of the subradiant state, which is much smaller than waveguide-induced dissipation. Therefore, efficient photon absorption can be realized by Bragg atom arrays without spoiling quantum coherence of atoms. To pinpoint the role played by subradiant state in photon absorption, we now analyze the role played by free-space dissipation in transmission and reflection processes. The CIT at resonance for the two-atom system without free space dissipation gives rise to
\begin{eqnarray}
\frac{\Xi_1}{\Gamma_1}+\frac{\Xi_2}{\Gamma_2}=0, \\
\frac{\tilde{\Xi}_1}{\Gamma_1}+\frac{\tilde{\Xi}_2}{\Gamma_2}=0.
\end{eqnarray}
Above equations indicate that subradiant and superradiant states produce destructive quantum interference in photon transmission and reflection processes. It is easy to see that the free-space dissipation does not affect interaction spectra $\Xi_j$ and $\tilde{\Xi}_j$ of scattering states. For the free-space dissipation which is equal to decay rate of the subradiant state, the scattering components for the superradiant state are almost unaffected. However, scattering components for the subradiant state become half of those without free-space dissipation. Accordingly, we obtain transmission and reflection amplitudes
\begin{eqnarray}
t &\approx& 1+\frac{\Xi_1 \Gamma}{2\Gamma_1}, \\
r &\approx& \frac{\tilde{\Xi}_1 \Gamma}{2\Gamma_1}.
\end{eqnarray}
It can be shown that $\Xi_1\Gamma/\Gamma_1=\tilde{\Xi}_1 \Gamma/\Gamma_1=-1$. Transmission and reflection amplitudes are $0.5$ and $-0.5$, respectively, giving rise to the photon absorption $\eta=0.5$. Therefore, photon absorption is enhanced by the subradiant state at weak free-space dissipation, i.e., collectively induced absorption (CIA). This effect makes it possible to realize photon detection without requiring large free-space dissipation~\cite{PhysRevLett.102.173602}, or topological protection~\cite{PhysRevApplied.15.044041}. In superconducting quantum circuits, the free-space dissipation is much smaller than waveguide-induced decay rate~\cite{mirhosseini2019cavity}. Therefore, the CIA effect we studied here is useful for photon detection in waveguide-coupled superconducting qubits.

Case (II) corresponds to the minimal absorption. The photon transmission is negligible because $\Gamma_f$ is larger than the linewidth of the transparency window. Therefore, the condition for CIT is not met. In Fig.~\ref{fig6}(d), we show reflectional components $\tilde{\Xi}/\mathrm{Im}(E)$ for the subradiant and superradiant states. When the free-space dissipation vanishes, these two collective states have destructive quantum interference in the reflection process. The component of the superradiant state is insensitive to the weak free-space dissipation. However, the subradiant state has decreasing contribution to photon reflection as $\Gamma_f$ increases. When the free-space dissipation is much larger than decay rate of the subradiant state, the subradiant state is ignorable for photon reflection. As a consequence, strong reflection is produced at $\Gamma_f=\delta/2$ where the superradiant state has a large value of $|\tilde{\Xi}/\mathrm{Im}(E)|$, and the component for the subradiant state is negligible.

Case (III) means that the free-space dissipation equals to decay rate of the superradiant state. Although photon absorption is the same for cases (I) and (III), their mechanisms are different. Case (III) is equivalent to the waveguide QED with a single atom, in which photon absorption can reach $0.5$ when the free-space dissipation and waveguide-induced decay rate are the same~\cite{PhysRevLett.102.173602}. In this scenario, the Bragg atom array behaves as a superatom where its optical response is solely determined by the superradiant state.

\begin{figure}[t]
\includegraphics[width=8.5cm]{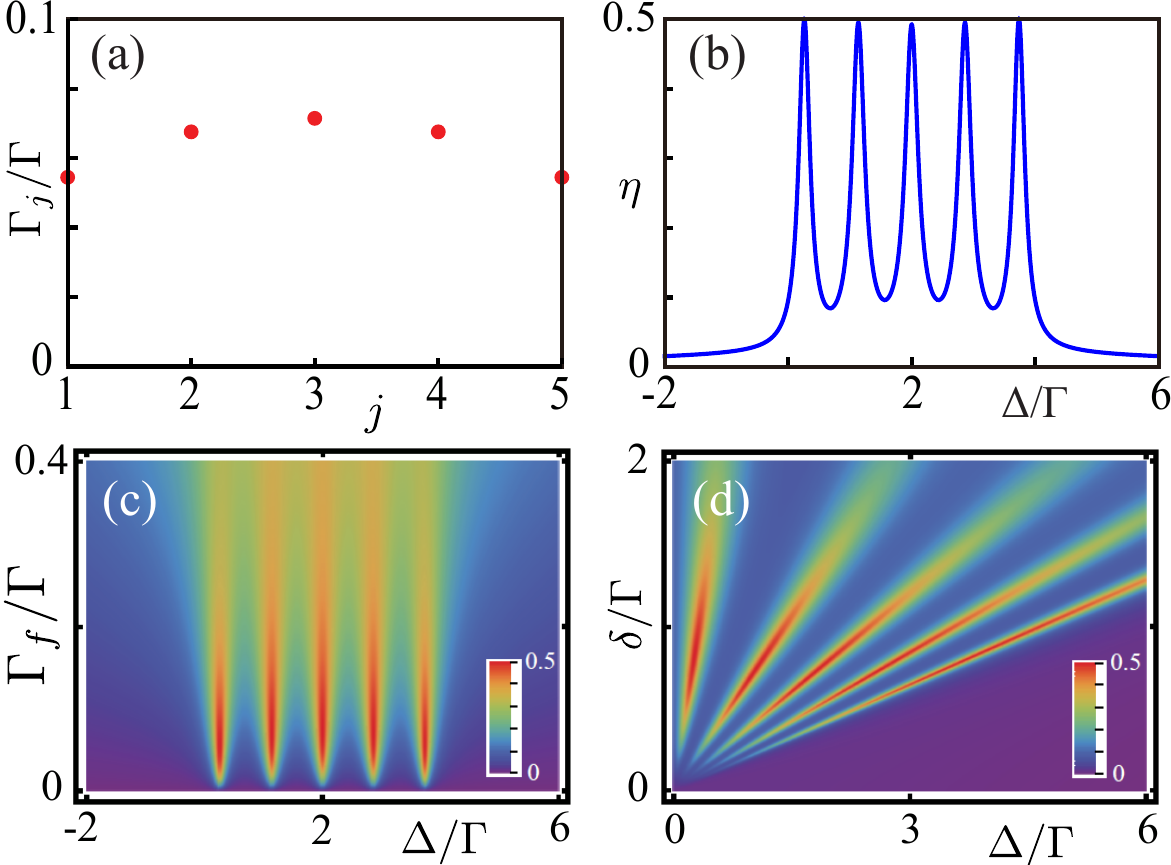}
\caption{(a) Decay rates of five subradiant states in the Bragg atom array with $N=6$. Here, $j$ labels the $j$-th subradiant state. (b) Photon absorption at different frequencies of subradiant states. (c) Photon absorption tuned by free-space dissipation. (d) Photon absorption tuned by equal difference $\delta$ in atomic frequencies. We consider the equal difference $\delta=0.8\Gamma$ in atomic frequencies in (a)-(c), and $\Gamma_f=0.055\Gamma$ in (b) and (d).}\label{fig7}
\end{figure}

Therefore, due to different decay rates of subradiant and superradiant states, the free-space dissipation plays an important role in manipulating quantum interference between these collective states in Bragg atom arrays. The CIA effect can be generalized to Bragg atom arrays with multiple transmission windows. In Fig.~\ref{fig7}(a), we show decay rates of subradiant states in the Bragg array with $N=6$ atoms. For simplicity, we assume that atoms in the array have frequency shifts $(0,\delta,2\delta,\ldots,5\delta)$ with equal difference $\delta$. In this scenario, subradiant states have decay rates close to each other. Hence, we can obtain multi-frequency CIA. In Fig.~\ref{fig7}(b), we show that photon absorption can be enhanced at different frequencies. These frequencies correspond to subradiant states in the Bragg atom array. We show multi-frequency photon absorption tuned by free-space dissipation in Fig.~\ref{fig7}(c). This multi-frequency CIA can be used for detecting multi-mode photons in waveguide QED systems. In experiment, the free-space dissipation is hard to be tuned. The condition for CIA can be met by modifying atomic frequencies. In Fig.~\ref{fig7}(d), photon absorption is nontrivially changed by the equal difference between atomic frequencies. Efficient photon absorption can be realized at different frequencies by employing multiple subradiant states.

\section{Summary}
We study collectively induced transparency and absorption in waveguide-coupled Bragg atom arrays. Atoms with inhomogeneous frequencies give rise to CIT. To pinpoint the role played by collective quantum states in CIT, we analytically study photon transport in a two-atom waveguide QED system. Destructive quantum interference between photons scattered by subradiant and superradiant states is responsible for the photon transparency. We further study Bragg arrays with many frequency-shifted atoms, and find multi-frequency photon transparency. Moreover, we discuss the effect of free-space dissipation. When free-space dissipation equals to the decay rate of a subradiant state, one can realize CIA. This enhanced photon absorption shows the significance of collective quantum states in controlling photon transport in a waveguide. Multi-frequency CIA is accessible when decay rates of subradiant states are close to each other. This can be obtained by considering atomic frequencies with equal difference. Our work presents a method for controlling subradiant states in waveguide-coupled Bragg atom arrays, and has potential applications in multi-frequency photon detection.

\begin{acknowledgments}
We thank Guo-Zhu Song for a critical reading. This work is supported by the National Natural Science Foundation of China (Grant No. 12105025).
\end{acknowledgments}

\end{document}